\documentclass[amsmath,amsfonts,aps,prd,showpacs,twocolumn,nofootinbib]{revtex4-1}
\pdfoutput=1
\usepackage[utf8]{inputenc}
\usepackage[T1]{fontenc}
\usepackage{lmodern,graphicx,cancel,multirow}

\usepackage[colorlinks=true, citecolor=blue, urlcolor=blue, linkcolor=blue, breaklinks=true, pdfpagelabels=false]{hyperref}

\begin{document}

\preprint{MADPH-11-1573, CP3-11-24}
\title{Baryon number violation at the LHC: the top option}
\author{Zhe Dong$^{a}$, Gauthier Durieux$^{b}$, Jean-Marc Gérard$^{b}$, Tao Han$^{a}$, Fabio Maltoni$^{b}$}
\affiliation{
$^{a}$Department of Physics, University of Wisconsin, Madison, WI 53706, USA\\
$^{b}$Centre for Cosmology, Particle Physics and Phenomenology (CP3), Université catholique de Louvain, Chemin du Cyclotron 2, B-1348 Louvain-la-Neuve, Belgium}

\begin{abstract}
Subject to strong experimental constraints at low energies, baryon number violation is nonetheless well motivated from a theoretical point of view. We examine the possibility of observing baryon-number-violating top-quark production or decay at hadron colliders. We adopt a model independent effective approach and focus on operators with minimal mass-dimension. Corresponding effective coefficients could be directly probed  at the Large Hadron Collider (LHC) already with an integrated luminosity of 1 fb$^{-1}$ at 7 TeV, and further constrained with  30 (100)  fb$^{-1}$ at 7 (14) TeV.
\end{abstract}
\pacs{12.38.Bx,12.60.-i,14.65.Ha,11.30.Fs}

\maketitle

\section{Introduction}

In the Standard Model (SM) for fundamental interactions, baryon ($B$) and lepton ($L$) numbers, associated with accidental global symmetries, are classically conserved quantities; a (tiny) violation is, however, induced by non-perturbative instanton effects~\cite{tHooft:1976rip}. Baryon number violation (BNV) also naturally occurs  in Supersymmetry~\cite{Weinberg:1981wj}, in Grand Unified Theories~\cite{Georgi:1974sy}, where BNV is notably mediated by new gauge bosons, and in black hole physics~\cite{Bekenstein:1971hc}. The cosmological production of matter from a matter--anti-matter symmetric initial condition moreover requires $B$ to have been violated in the early Universe~\cite{Sakharov:1967dj}. 

On the other hand,  experimental constraints on several BNV processes have reached impressive heights. Nucleon decay channels provide the best examples, though baryon-number-violating decays of the $\tau$ lepton or, much more recently, of heavy mesons have also been investigated~\cite{ParticleDataGroup:2010dbb}. These latter measurements opened the way for direct experimental tests of the baryon number conservation law within the second and third generations of quarks and leptons  but have not  really extended the range of energy scales. The only direct experimental constraints on BNV beyond the GeV scale are bounds on the $Z\to p\, e^-,\: p\,\mu^-$ branching ratios obtained at LEP~\cite{ParticleDataGroup:2010dbb}. The LHC comes as a natural step forward in probing  the baryon number conservation law beyond the TeV scale and the first generation (see, {\it e.g.}, Ref.~\cite{Morrissey:2005uza}). In particular, with a very large production rate and unique experimental signatures,  top quarks are an interesting option: the top flavor can be clearly identified, the $t$ and $\bar t$ are distinguishable via the charged lepton in their decay, and the hadronization effects unimportant. Consequently,  BNV could be probed at the quark level.

We choose to consider interactions involving one single top quark and a charged lepton.  The presence of a single final state lepton produced from the proton-proton initial state implies a total change in lepton number $\Delta L = \pm 1$, and, by conservation of angular momentum $\Delta (L+3B)\in 2\mathbb{Z}$, requires a simultaneous violation of $B$. 
Thus a single charged lepton without missing energy  points toward BNV. In the presence of neutrinos
 in the production process, though,  the lepton number is intractable. Consequently,  baryon-number-violating processes would be more difficult to identify unambiguously, and could, for instance, be confused with  flavor-changing neutral currents~\cite{Andrea:2011ws,Kamenik:2011nb}.

\section{Effective operators}

The effective BNV Lagrangian  can easily be built out of five lowest dimensional  effective operators~\cite{Weinberg:1979sa,Wilczek:1979hc,Abbott:1980zj}  that preserve Lorentz invariance and $SU(3)_C\otimes SU(2)_L\otimes U(1)_Y$ gauge symmetries, along with an accidental global $B-L$ symmetry. Following the notation of Ref.~\cite{Weinberg:1979sa}, we write
\begin{equation}
{\cal L}^{\rm dim=6}_{\rm BNV} = \frac{1}{\Lambda^2} \sum_{i=1}^{5} c_i \,O^{(i)}  \,,
\label{bnv}
\end{equation}
where $c_i$ are the effective (dimensionless) coefficients of the corresponding operators $O^{(i)}$ and $\Lambda$ is the mass scale associated with physics responsible for BNV beyond the Standard Model. 

Expanding $SU(2)_L$ indices in the operators $O^{(1-5)}$ and identifying one up-type quark as the top,  the effective terms that do not contain neutrinos can be parametrized as linear combinations of only two operators (and their Hermitian conjugates),
\begin{equation}\hspace{-0mm}
\begin{aligned}
&O^{(s)} \equiv \epsilon^{\alpha\beta\gamma}
[ \overline{t^c_\alpha} (a P_L + b P_R ) D_\beta	] 
[ \overline{U^c_\gamma} (c P_L + d P_R) E ] 
,\\
&O^{(t)} \equiv \epsilon^{\alpha\beta\gamma}
[\overline{t^c_\alpha} (a' P_L + b' P_R ) E	 ]
[\overline{U^c_\beta}  (c' P_L + d' P_R) D_\gamma ],
\end{aligned}\hspace{-3mm}
\label{st-basis}
\end{equation}
where $D,\, U,\, E$ respectively denote generic down-, up-type quarks and charged leptons. We emphasize that fermions  in Eq.~\eqref{st-basis}  are taken as mass eigenstates. Charge conjugated fields are defined as $\psi^c\equiv C\overline{\psi}^T$ with $C$, the charge conjugation matrix;  $2P_{L/R}\equiv 1\mp \gamma^5$; colors are labeled by Greek indices;  $a,\ a',\ldots$ are fermion-flavor-dependent  effective parameters.

Three comments are in order. First, the $(s)$, $(t)$ labeling in Eq.~\eqref{st-basis} reminds that the scale $\Lambda$ in Eq.~\eqref{bnv} may be linked to the mass of a heavy mediator (with electric charge $1/3$) exchanged in $s$ or $t$ channels, respectively. If so identified, then the coupling parameters $a,\ a',\ldots$ could be naturally of the order of unity.
We stress that the operator  $O^{(u)}\!\equiv\epsilon^{\alpha\beta\gamma} \;[\overline{t^c_\alpha} P_L U_\beta]  [\overline{D^c_\gamma} P_L E$] arising from $O^{(4)}$ and possibly associated with a mediator of electric charge $4/3$ exchanged in the $u$ channel does not need to be introduced at the effective level.  The reason is that the Schouten identity,
\begin{gather*}
[CP_L]_{ij}[CP_L]_{kl}-[CP_L]_{ik}[CP_L]_{jl}
+ [CP_L]_{il}[CP_L]_{jk}=0,
\end{gather*}
 can be used to express $O^{(u)}$ in terms of $O^{(s)}$ and $O^{(t)}$, {\it i.e.}, $O^{(u)}=O^{(s)} (a=c=1,\, b=d=0) - O^{(t)} (a'=c'=1,\, b'=d'=0)$. Second, heavy gauge mediators (vectors) give rise to $O^{(1,2)}$ only~\cite{Weinberg:1979sa,Dorsner:2004xa}, which in our basis, Eq.~\eqref{st-basis}, entails $a=0=d$ or $b=0=c$ (or primed analogs). Third, operators involving two top quarks can also be obtained from Eq.~\eqref{st-basis} by substituting $t$ for $U$. Note, however, that in this case $O^{(s)}$ and $O^{(t)}$ are no longer independent and considering only one of the two is then sufficient. Such operators could, for example, mediate processes like $e^- d\to \bar{t}\,\bar{t} $ in future $e^-p$ colliders, or $gd\to  \bar{t}\,\bar{t} e^{+}$ at the LHC.
 
\section{Processes} 

At the LHC, possibly relevant BNV processes involving a top quark are
\begin{equation}
\renewcommand{\arraystretch}{1.2}
\begin{array}{ll}
t \xrightarrow{\text{BNV}}\overline{U}\,\overline{D}\, E^+	&({\rm decay})\\
U\, D  \xrightarrow{\text{BNV}} \bar{t}\, E^+ &({\rm production})
\end{array}
\label{processes}
\end{equation}
(and their charge conjugate analogs) where, in the first case,  top quarks are produced through SM processes. Since, as mentioned above, a single charged lepton without any missing transverse energy ($\cancel{E}_T$) in the final state is a clear signal for BNV, it is simpler to avoid signatures that lead to neutrinos in the final state. Fully reconstructed top leptonic decays could be considered in more refined analyses.
We  also note that the flavor assignments can be very relevant from the phenomenological point of view. In decay,  heavy flavors such as charm and bottom could be tagged in jets.  In production, the relevance of initial quark flavors is determined also by proton parton distribution functions (PDFs).

Neglecting all fermion masses but the top one, $m_t$, and using the algebraic rules introduced in Ref.~\cite{Denner:1992vza}, the squared amplitude for the processes in Eq.~\eqref{processes} induced by the operators of Eq.~\eqref{st-basis}  reads
\begin{align}
\sum_\text{\scriptsize\parbox{1cm}{\centering spins, colors}} |\mathcal{M}|^2
= \frac{24}{\Lambda^4}\Big[
 &(p_t\cdot p_D)\: (p_U\cdot p_E)\: (A+C)\nonumber\\[-4.5mm]
-&(p_t\cdot p_U)\: (p_D\cdot p_E)\: C\nonumber\\
+&(p_t\cdot p_E)\: (p_D\cdot p_U)\: (B+C)
\Big]\,.
\end{align}
The dimensionless parameters 
\begin{equation}
\begin{aligned}
A&\equiv \left(|a|^2 +|b|^2\right)\left(|c|^2+|d|^2\right),\\
B&\equiv \left(|a'|^2 +|b'|^2\right)\left(|c'|^2+|d'|^2\right),\\
C&\equiv \mathfrak{Re}\big\{ a^*c^*a'c'+ b^*d^*b'd'\big\},
\end{aligned}
\label{abc}
\end{equation}
arise respectively from the square of $O^{(s)}$, of $O^{(t)}$ and from the interference between these two operators (which is absent if BNV is mediated by vectors only).
\begin{figure}[b]
  \centering
  \includegraphics[width=0.92\columnwidth]{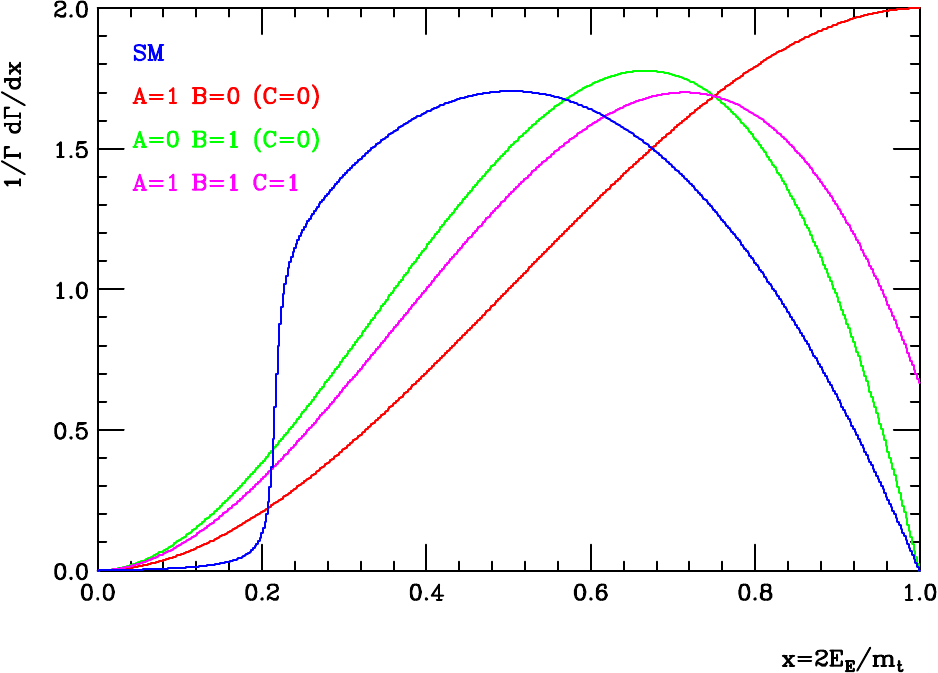}
  \caption{Energy spectrum of the charged lepton in the SM  $t \to b E^+ \nu_E$ (blue curve) and in BNV  $t  \to \overline{U}\, \overline{D}\, E^+$ top decays. }
  \label{fig:1}
\end{figure}

For a BNV decay, we obtain the following partial width:
\begin{align*}
\Gamma^\text{BNV}_t\! &=\hspace{-2.3mm}\int\limits_{0}^{m_t/2}\hspace{-2mm}\text{d}E_{\scriptscriptstyle E}\:
\frac{m_t^2E_{\scriptscriptstyle E}^2}{32\pi^3\Lambda^4} \left[\!\left(\frac{A}{3}+B+C\right) \left(
1- \frac{2E_{\scriptscriptstyle E}}{m_t} \right) + \frac{A}{6} \!\right]\\
&= \frac{m_t^5}{192\pi^3}\frac{1}{16\Lambda^4}\left[A+B+C\right]\,,
\end{align*}
where $E_{\scriptscriptstyle E}$ is the lepton energy in the top rest frame. In Fig.~\ref{fig:1}, we compare the charged lepton energy spectrum
in a SM decay to that in a BNV decay for three different representative choices of $A, B$ and $C$. 
 Inputing the SM width for the top quark ($1.4$~GeV), the BNV branching ratio can be conveniently written as
\begin{align*}
&{\rm Br}^\text{BNV}_t\!= 1.2 \times 10^{-6}
\left(\frac{m_t}{173~\text{GeV}}\right)^5
\left(\frac{1\text{TeV}}{\Lambda}\right)^4
\left[A+B+C\right].
\end{align*}
Taking  the  $t\bar{t}$  production cross section  at the 7 (14) TeV LHC to be  $150$ (950)~pb, we can expect
0.35 (2.2)$/$fb$^{-1}$ BNV top decays if $A+B+C=1$, for each allowed flavor combination.

For BNV production, the partonic cross section reads
\begin{align}
\hat{\sigma}^\text{BNV}_t\! &= \frac{1}{96\pi\Lambda^4}\int\limits_{m_t^2-\hat{s}}^0\hspace{-3mm}\text{d}\hat{t}\:
\left[ A \frac{\hat{t}\left(\hat{t}-m_t^2\right)}{\hat{s}^2}
+ B \frac{\left(\hat{s}-m_t^2\right)}{\hat{s}} \nonumber
- 2C \frac{\hat{t}}{\hat{s}}\,\right]\\
&= \frac{\hat{s}}{96\pi\Lambda^4} 
\left(1-\frac{m_t^2}{\hat{s}}\right)^2
\left[\left(\frac{A}{3}+B+C\right)+\frac{m_t^2}{\hat{s}}\frac{A}{6}\right]\,,
\label{cross-section}
\end{align}
with the Mandelstam variables $\hat{s}\equiv (p_U+p_D)^2$ and $\hat{t}\equiv(p_U-p_E)^2$.  
As expected from dimensional arguments the cross section induced by the operators in Eq.~\eqref{st-basis} grows as $\hat{s}/\Lambda^4$. However, in setting lower bounds on the scale of new physics, it is important to always keep in mind that  the validity (and unitarity) of the effective field theory itself assumes $\hat{s} \ll \Lambda^2$. 
 
Out of the six possible initial quark flavor assignments, (namely, $ud,\,us,\, ub,\,cd,\, cs$ and $cb$), we consider  
\begin{equation*}
\renewcommand{\arraystretch}{1.5}
\begin{array}{l@{\hspace{4mm}-\hspace{1.5mm}}p{4.8cm}}
u\,d\to \bar{t}\, E^+  &the most PDF-favored,	 \\
u\,b\to \bar{t}\, e^+		&possibly flavor-unsuppressed,\\
c\,b\to \bar{t}\, \mu^+	&the most PDF-suppressed, yet,\\[-2mm]
&possibly flavor-unsuppressed,
\end{array}
\end{equation*}
as well as  their charge conjugate analogs. Operators with two pairs of fermions in the same generation could be favored by the flavor structure of the underlying theory.  In Table~\ref{tab:1}, we collect the cross sections for the different processes at the LHC with $\sqrt{s}=7$ (14)  TeV. To enforce unitarity in a simple yet efficient and model-independent way, we impose $\sqrt{\hat{s}}<\Lambda$. This cut has an important effect on valence quark initiated processes but a very mild one on processes initiated by sea quarks. 

\begin{table}
\begin{tabular}{c|*{3}{@{\hspace{6mm}}c}}
$\sigma\!\,$[fb]
&$ud\to \bar{t} E^+$& $ub\to \bar{t}e^+$&$cb\to\bar{t}\mu^+$    \\
$A \; B \; C $
&$\bar{u}\bar{d}\to t E^- $& $\bar{u}\bar{b}\to t e^- $& $\bar{c}\bar{b}\to t \mu^-$\\
\hline\hline
\multirow{2}{*}{$1\;\;  0 \;\; 0$}
&	250 (690)	&	30 (150)	&	1.2 (10) \\
&	14 (74)             &      3.1 (21)     &       1.2 (10) \\\hline
\multirow{2}{*}{$0\;\; 1\;\; 0$} 
&	910	 (1\,900) &	110	(440) &	3.7 (28) \\
&	45	(220)        &	9.1	(60)   &	3.7 (28) \\\hline
\multirow{2}{*}{$1\;\; 1\;\; 1$}
&	2\,100 (4\,600)	&	240 (980)	&	9.1 (66) \\
&	110	(500)                   &	22 (140)	         &	9.1 (66)
\end{tabular}
\caption{Cross sections (fb) for representative BNV production processes at the LHC, with three different choices of 
$A,B$ and $C$,  
$\sqrt{\hat{s}}<\Lambda=1$~TeV, 
$\sqrt{s}=7$~TeV (14 TeV in parentheses) and CTEQ6L1 PDF~\cite{Pumplin:2002vw} (renormalization and factorization scales set at $m_t=173$~GeV).}
\label{tab:1}
\end{table}

\section{LHC searches} 

We now briefly discuss BNV signatures at the LHC.  For the sake of illustration we make a definite choice for the fermion flavors  in Eq.~\eqref{processes} and consider\\
I) \underline{BNV decay}: $pp \xrightarrow{\text{SM}} t\, \bar{t}$ with the top decaying via a BNV interaction $t \xrightarrow{\text{BNV}} \bar b\, \bar c\, \mu^+$  and the anti-top decaying fully hadronically, which leads to the  $\mu^+$+5-jet final state; 
II) \underline{BNV production}:  $p\,p \xrightarrow{\text{BNV}} \bar t\,  \mu^+$  with $u,d$ flavors in the initial state, the anti-top decaying fully hadronically, leading to $\mu^+ +3$~jets. 

The first interesting observation is that there are no irreducible backgrounds to such signatures as both of them have no $\cancel{E}_T$.   On the other hand, processes resulting from a leptonically decaying $W$ with a small reconstructed  $\cancel{E}_T$ could mimic the signal. A proper investigation of such backgrounds requires not only parton showering, hadronization and realistic detector simulation but also data driven methods.  However, a few relevant observations can already be made with a simple parton-level simulation. To this aim, we have implemented BNV interactions in \textsc{MadGraph~5}~\cite{Alwall:2011uj}  via \textsc{FeynRules}~\cite{Christensen:2008py} and generated events for both signal and representative backgrounds in the same simulation framework.

The search for BNV decays proceeds through the selection of $\mu^+ +5$~jets with an upper cut on the $\cancel{E}_T$. The presence
of two tops, one hadronically decaying $W$ and possibly  two $b$-tagged jets can be efficiently used to better reconstruct the event  kinematics. In addition, note  that the BNV decay of a top  quark gives $\mu^+ \bar b$ at variance with the SM semi-leptonic decay which gives $\mu^+ b$.   Determining the bottom quark  charge ({\it e.g.}, via a lepton tag) in the BNV decay could therefore offer crucial discrimination power. The main SM backgrounds to this signature come from  $t\, \bar{t} +1$~jet  and $W^++5$~jets, the former being dominant after $b$ tagging.  
\begin{figure}[b]
  \centering
  \includegraphics[width=0.92\columnwidth]{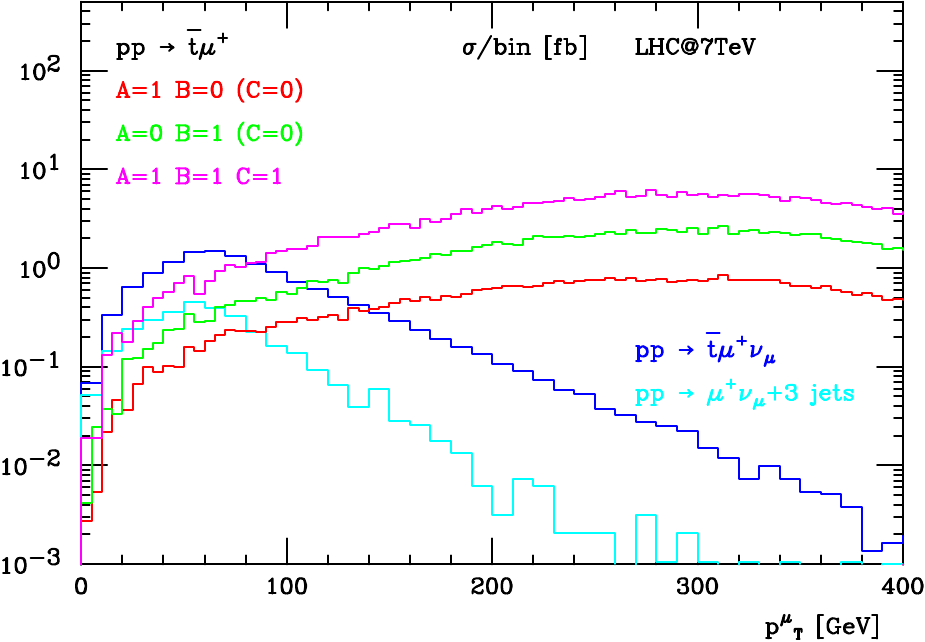}
  \caption{Transverse momentum for the charged lepton in the BNV production signal $\bar t\,  \mu^+$ (from $ud$ initial state) and in the $W^++$3-jet and $\bar t W^+$ backgrounds. Top quarks are decayed hadronically. Selection cuts on the three jets and the muon are given in the text.}
  \label{fig:2}
\end{figure}

The search for BNV production proceeds through the selection of $\mu^+ +3$~jets with an upper cut on the $\cancel{E}_T$. The reconstruction is simpler than in the BNV decay search, as there is no combinatorial background and the top and $W$ mass constraints can be used to improve the resolution on the signal kinematics.  In Fig.~\ref{fig:2}, we compare the $p_T$ of the charged lepton in the signal to that of the $W^+$+3-jet and $\bar{t}\: W^+$ (with $W^+ \to \mu^+ \nu_\mu$) backgrounds. We require three central jets ($p_T>40$~GeV$,|\eta|<2.5,\Delta R_{jj}>0.5$), a central isolated lepton ($|\eta|<2.5,\Delta R_{j\mu}>0.5$) and  $\cancel{E}_T<30$ GeV. In the $W^+$+3-jet  background we also demand $|m_{jjj} - m_t|<40$ GeV and a $b$ tag. As expected from the $\hat{s}$ enhancement in the cross section, tamed by requiring $\sqrt{\hat{s}}<\Lambda=1$~TeV, the $p_T$ distribution of the lepton in the signal is much harder than in the backgrounds.

The LHC reach for the processes in Table~\ref{tab:1} can be expressed in terms of  the minimal value for the parameters defined in  Eq.~\eqref{abc} leading to a sensitivity $S/\sqrt{S+B}\ge 5$.  For the sake of illustration we consider 
30 (100) fb$^{-1}$ of collected luminosity at the LHC for $\sqrt{s}=7$ (14) TeV, 
 the event selection described in the above paragraph with the additional requirement $p_T>150$~GeV for the charged lepton (one flavor), 
 both $t$ and $\bar t$ production,  
only the $tW$ background,
and  $A=B=C$. In so doing, we find 
\begin{equation*}
\renewcommand{\arraystretch}{1.0}
\begin{array}{l@{\hspace{1.5mm}:\hspace{4mm}}p{3cm}p{2cm}}
u\,d\to t\, E       & $A,\,B,\,C \ge 0.0076$   & (0.0046) \\
u\,b\to t\, e	    & $A,\,B,\,C \ge 0.084$  &(0.026)	\\
c\,b\to t\, \mu 	& $A,\,B,\,C \ge  1.6$   & (0.21)
\end{array}
\label{procs}
\end{equation*}
which point to a sensitivity at the $10^{-1}-10^{-2}$ level for the effective coefficients $c_i$ of Eq.~\eqref{bnv} at the TeV scale.

Finally, we stress that, for both BNV production and decay signatures, selecting high-$p_T$ tops could be advantageous. In  this limit, the BNV production signal is enhanced with respect to the backgrounds, while for the BNV decay search, top decay products might cluster into one jet, curbing, for instance, the combinatorial problems in the $\mu^+$+5-jet signature and also controlling better $\cancel{E}_T$ uncertainties. To this aim, efficient boosted reconstruction techniques for the top quark should be  employed~\cite{Plehn:2009rk}.

\begin{figure}[t]
\mbox{\hspace{-1mm}
\includegraphics[scale=.56]{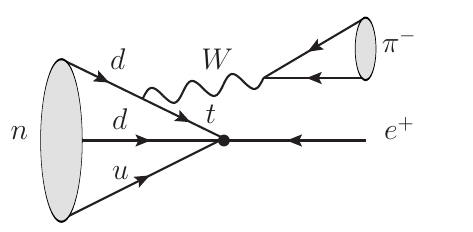}
\includegraphics[scale=.56]{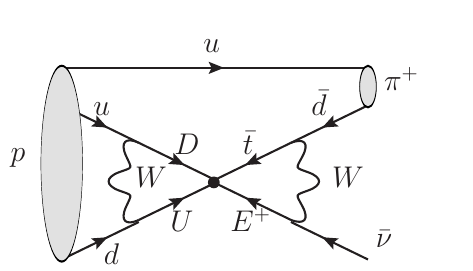}}\\
(a)\hspace{4cm} (b)
\caption{Representative (a) tree-level and (b) two-loop-level diagrams involving the BNV operators given in Eq.~\eqref{st-basis}
and leading, in principle, to nucleon decay.}
\label{fig3}
\end{figure}

\section{Indirect constraints}

We finally consider how direct limits on BNV interactions that can be extracted at colliders compare with the formidable ones available from nucleon decays.
For example, it is straightforward to see that  the LEP bound on BNV interactions in $Z\to p e^-$  quoted in the introduction is orders of magnitude weaker than the corresponding one obtained by the constraints on proton decay via $p \to e^+Z^*\to e^+e^-e^+$.
Operators  considered in Eq.~\eqref{st-basis}  also contribute indirectly to nucleon decays~\cite{Hou:2005iu} through tree and/or loop diagrams.
Tree-level diagrams with one $W$-emission, such as  that in Fig.~\ref{fig3}a provide formidable high lower bounds on $\Lambda$ (or equivalently upper bounds on the effective parameters) if the lepton is not a $\tau$. In fact, two $W$-emissions are needed for a $u d t\tau$-operator to be relevant in nucleon decays and the constraints become weaker.
Moreover, if the dominant BNV dimension-six operators only involve the third and second generations of quarks and leptons,  three $W$-emissions are required, and the rate is suppressed to a level consistent with the data.
In BNV production, these theoretical considerations tend to favor the PDF-suppressed processes  of Table~\ref{tab:1}.
By considering a single operator contribution at a time, with fixed flavors in the two-loop diagram of Fig.~\ref{fig3}b, extremely small upper bounds on the effective parameters can also be obtained~\cite{Hou:2005iu}. Yet, strong cancellations may occur when summing over all possible $UDUE$ virtual contributions and allow effective parameters to be large (say, of order one). Mechanisms that could lead to such \emph{GIM-like} cancellation at one- and two-loop level remain to be examined within a complete theory for flavor, starting with dimension-six BNV operators expressed in terms of weak eigenstates. While efforts in this direction are on going, we adopt a pragmatic attitude and encourage the LHC experimental collaborations  to set genuine and direct bounds on BNV without any theoretical prejudice.\vspace*{-.1cm}

\section{Conclusions}

We have studied lowest dimensional BNV operators and their consequences for top physics at the LHC. Corresponding effective coefficients could be probed directly at the TeV scale up to the $10^{-1} - 10^{-2}$ level. In this prospect, possible flavor models ({\it e.g.}, warped extra-dimensions or horizontal symmetries) leading to  nucleon decay rates consistent with the present experimental bounds should be examined.

\section*{Acknowledgements}

We thank Giacomo Bruno, Michele Gabusi and Davide Pagano for very enjoyable and stimulating discussions on promising BNV signatures at the LHC and Christopher Smith and Pavel Fileviez Perez for insights on  the  flavor aspects of nucleon decays. This research has been  supported in part by the U.S.~Department of Energy under grant No.~DE-FG02-95ER40896, by the Belgian IAP Program BELSPO P6/11-P and by the IISN.

\bibliographystyle{apsrev4-1_title}
\bibliography{bnv}
\end{document}